# Comment on "Significant-Loophole-Free Test of Bell's Theorem with Entangled Photons"

Louis Vervoort, 04.02.2016

*Institut National de Recherche Scientifique, Montreal, Canada*
*Minkowski Institute, Montreal, Canada*

**Abstract**. In a recent article Giustina et al. [Phys. Rev. Lett. 115, 250401 (2015)] report on an advanced Bell experiment, simultaneously closing loopholes for local hidden-variable theories. The authors claim that 'local realism' has been refuted, unless 'truly exotic hypotheses' are made. Here I argue that a particularly wide and natural class of local hidden-variable theories survives, for instance when the hidden variables describe a background field. Such background-based theories exploit the freedom-of-choice loophole, which cannot be closed for this type of hidden variables. The dynamics of such models can be illustrated by existing systems, e.g. from fluid mechanics.

In a recent article [1], a team of physicists led by Anton Zeilinger reports on a remarkable Bell experiment, simultaneously closing the most relevant loopholes for local hidden-variable theories (HVTs). Almost at the same time a second [2] and third [3] article have been published by other teams, drawing largely the same conclusions. In these experiments the so-called loopholes are closed by a sophisticated experimental set-up ensuring that, for each particle pair, the relevant emission event, measurement events, and setting choice events are mutually spacelike separated. Upon closer inspection, the authors of Ref. [1] appear to claim two essential results, which may seem equivalent to the non-expert, but which are actually quite different. The aim of this Comment is, first, to highlight the difference between both claims in greater detail than is done in [1], and second to argue that a wide and natural class of local HVTs still survives. This seems essential in view of the far-reaching consequences these experiments may have for theoretical physics. Here I focus on Ref. [1], since it is more explicit about the assumptions made, but similar remarks hold for Refs. [2-3].

The authors' first claim (Claim 1) is very broad, and asserts that 'local realism' has been refuted by their experiment, unless some 'truly exotic hypotheses' are made. 'Realism' is 'the assumption that objects have physical properties independent of measurement', and 'locality' is the assumption that there are no influences traveling faster than the velocity of light [1]. So it is claimed that at least one of these cornerstones of physics must be given up: 'Our experiment provides the strongest support to date for the viewpoint that local realism is untenable.' The second, more specific claim (Claim 2) – which can indeed, I believe, legitimately be inferred from the experiment –, amounts to the assertion that a *certain class* of local HVTs has been refuted by the experiment. To their credit, the authors of [1] explicitly specify the content of Claim 2. Thus on page 2 the authors describe how the experiment would close the 'freedom-of-choice' loophole (the crucial passage is italicized): 'The freedom-of-choice loophole refers to the requirement, formulated by Bell, that the setting choices are "free or random" […]. For instance, this would prohibit a possible interdependence between the choice of measurement settings and the properties of the system being measured. Following Bell, we describe all properties of the system with the variable $\lambda$, which represents "any number of hypothetical additional complementary variables needed to complete quantum mechanics in the way envisaged by EPR" […]. *This loophole can be closed only under specific assumptions about the origin of $\lambda$. Under the assumption that $\lambda$ is created with the particles to be measured*, an experiment in which the settings are generated independently at the measurement stations and spacelike separated from the creation of the particles closes the loophole.' So, according to Claim 2 the experiment has ruled out, with overwhelming likelihood, all local HVTs in which the HVs $\lambda$ describe intrinsic properties of the particle pair that are created at the source.

Now, when considering the possibility of completing the quantum prediction of the EPR and Bell experiments, it is certainly very natural to consider the $\lambda$ as 'pertaining' only to the particle pair, more generally as created at the source with the particles, in the sense of [1]. But it is not the only possibility that is logically allowed. The point of this Comment is that the class of HVTs still exploiting the freedom-of-choice loophole may be vast and natural. For instance, the $\lambda$ could, besides the particles, also describe a background field or medium in which the particles move and that interacts with particles and analyzers [4, 8]. In this case the full dynamics of the $\lambda$ is crucial, as was recently investigated in detail in Ref. [4] (the background field may refer to vacuum fluctuations, an ether, a dark field,…). Specifically, let $\lambda \equiv (\lambda_0, \lambda_1, \lambda_2)$, where $\lambda_0$ are properties belonging to the particle pair in the sense of [1], $\lambda_1$ properties of the background field in the neighborhood of the left analyzer (at angle a), and $\lambda_2$ properties of the field close to the right analyzer (b). Then it is clear that the conditional probability $P(\lambda|a,b) \equiv P(\lambda_0, \lambda_1, \lambda_2|a,b)$ is in general different from the unconditional $P(\lambda_0, \lambda_1, \lambda_2)$ simply because $\lambda_1$ can interact with analyzer 'a' and $\lambda_2$ with analyzer 'b'; therefore $\lambda$ can obviously be dependent on (a, b) even if the interactions are entirely local. In such a 'background-based' HVT there *is* a probabilistic dependence between the λ and (a,b), but this in no way means that the settings (a,b) are conspiratorially determined by the λ; there is just a stochastic correlation, as happens in countless probabilistic experiments [4]. That such background-based HVTs can reproduce the quantum correlation of the Bell experiment is shown in [4]. To do so one needs to assume a specific dynamics between the particles ($\lambda_0$), the analyzers (a, b) and the background field ($\lambda_1, \lambda_2$); but such a dynamics is a logical possibility and not precluded by any physical law. That the freedom-of-choice loophole is more subtle than meets the eye, has recently been argued by several researchers [8-11]. What is important for this Comment, independent of the details and even of the validity of the case study [4], is that there may be HVTs that do *not* assume 'that λ is created with the particles to be measured'. In our example $\lambda_1$ and $\lambda_2$ are field values of a background in the spacetime region of the measurement, not emission, events.

To give further physical grounds why in particular background-based HVTs are relevant in this discussion, it suffices to have a look at the spectacular experiments recently performed by Couder, Fort, Bush and coworkers [5-6]. These researchers have shown that oil droplets can, under specific conditions, be made to walk over a vibrating fluid film, and mimic a wide range of quantum phenomena, including double-slit interference, orbit quantization and Zeeman splitting. The essential 'element of reality' that is responsible for such quantum-like behavior is the pilot or background wave that accompanies the droplets. The droplets hit the fluid film and create a surface wave on it, which guides their movement. In such fluid mechanical systems there is a complex dynamics between the (properties of the) droplets (the equivalent of $\lambda_0$ above), the background or pilot wave ($\lambda_1, \lambda_2$), and the detailed geometry of the fluid bath or any 'contextual' variables (a, b). Such systems exhibit massive correlations [6]. There are e.g. manifestly correlations of the type $P(\lambda_1, \lambda_2|a,b)$: the pilot wave is strongly dependent on, e.g., the geometry of the bath (yet (a,b) can be free or random variables). The probabilistic dependencies one has to assume in the background-based model for the Bell experiment [4], are of the same type as exist in the droplet systems.

Of course, the background-based theories we have in mind do remind one of Louis de Broglie's pilot-wave theory. There is a whole series of physicists working on modern variants of this theory, attempting to derive quantum mechanics from a hidden level of reality (see a condensed review in [6]). One can also refer here to the cellular automaton theory of quantum mechanics of 't Hooft, which indeed features correlations as above [7].

In conclusion, the crucial experiment by Giustina et al. does eliminate a wide and natural class of local-realistic models (Claim 2). But when more general claims are made ('By closing the freedom-of-choice loophole […] we reduce the possible local-realist explanations to truly exotic hypotheses'), we disagree. Essentially the same remark holds, mutatis mutandis, for Refs. [2-3]. In view of the importance of these experiments for physics – after all, the unification of quantum mechanics and relativity theory may be at stake – it seems important to highlight the distinction.

*Acknowledgements. I thank Ronald Hanson for his willingness to discuss his group's experiment and its interpretation.*

[1] M. Giustina et al., 'Significant-loophole-free test of Bell's theorem with entangled photons', Phys. Rev. Lett. **115**, 250401 (2015)

[2] L. Shalm et al., 'Strong loophole-free test of local realism', Phys. Rev. Lett. **115**, 250402 (2015)

[3] B. Hensen et al., 'Loophole-free Bell inequality violation using electron spins separated by 1.3 kilometers', Nature **526**, 682 (2015)

[4] L. Vervoort, 'No-go theorems face background-based theories for quantum mechanics', Found. Physics **45**, 1 (2015), doi: 10.1007/s10701-015-9973-7

[5] A. Eddi, J. Moukhtar, S. Perrard, E. Fort, and Y. Couder, 'Level-Splitting at a macroscopic scale', Phys. Rev. Lett. **108**, 264503 (2012)

[6] J. Bush, 'The new wave of pilot-wave theory', Phys. Today, Aug. 2015, 47

[7] G. 't Hooft, 'Models on the boundary between classical and quantum mechanics', Phil. Trans. R. Soc. **A 373**, 20140236 (2015)

[8] A. Khrennikov, 'Prequantum classical statistical field theory: background field as a source of everything ?', J. Phys.: Conf. Ser. **306**, 012021 (2011)

[9] A. Khrennikov, *Interpretations of Probability*, de Gruyter, Berlin (2008)

[10] M.J.W. Hall, 'Local deterministic model of singlet state correlations based on relaxing measurement independence', Phys. Rev. Lett. **105**, 250404 (2010)

[11] M.J.W. Hall, 'Relaxed Bell inequalities and Kochen-Specker theorems', Phys. Rev. A **84**, 022102 (2011)